# Relativistic length expansion in general accelerated system revisited.

J.Foukzon[1], S.A. Podosenov[2], A.A.Potapov[3].

[1] Israel Institute of Technology.
[2] All-Russian Scientific-Research Institute
 for Optical and Physical Measurements,
[3] IRE RAS,

**Abstract**. The aim of the present article is to give an exact and correct representation of the essentially important part of modern special relativity theory that touches upon the behavior of the proper length of accelerated moving bodies. In particular we pointed out that standard solution of the Bell's problem [3]-[4] revision needed. Classical solution of the relativistic length expansion in general accelerated system completely revisited. Instant proper length measurement between J.S.Bell's rockets also is considered successfully.

## 1. Introduction

As known [1],[2] constructing the covariant SRT on a general non-inertial frame one should exactly distinguish a coordinates (in some sense formal-mathematical) of a particles and *physical distance* (experimentally measurable) one. The latter is defined as the ratio of the *physical distance* $dl_{\mathbf{ph}}$ and *physical time* $d\tau_{\mathbf{ph}}$ :

$$ds^2 = c^2 d\tau_{\mathbf{ph}}^2 - dl_{\mathbf{ph}}^2, \tag{1}$$

$$dl_{\mathbf{ph}}^2 = \left(-g_{\alpha\beta} + \frac{g_{0\alpha}g_{0\beta}}{g_{00}}\right)dx^\alpha dx^\beta, \qquad (2)$$

$$d\tau_{\mathbf{ph}} = \sqrt{g_{00}}\,dt + \frac{g_{0\alpha}dx^\alpha}{c\sqrt{g_{00}}}. \qquad (3)$$

Many authors use Eq.(2) directly. In particular for relativistic length expansion for homogeneously accelerated rod [3] by using Eq.(2) one obtain

$$dl_{\mathbf{ph}} = l_0\sqrt{1 + \left(\frac{at}{c}\right)^2}. \qquad (4)$$

However, the infinite small *inertial* rod proper length in SRT is defined by the space part of the (complete space-like) 4-vector

$$dl_{\mathbf{i}} = (cdt, dx, 0, 0), dt = 0. \qquad (5)$$

For simplicity, the rod is oriented and moves along the X-axis, therefore $dy = dz = 0$ and $dx$ is the rod length $dl_{\mathbf{i}}^{pr}$. Hence in non inertial case, i.e. for the infinite small non inertial rod proper length one obtain the next definition

$$dl_{\mathbf{a}}^{pr} = (cd\tau_{\mathbf{ph}}, dl_{\mathbf{ph}}, 0, 0),$$

$$d\tau_{\mathbf{ph}} = 0 \qquad (6)$$

where

$$d\tau_{\text{ph}} = \sqrt{g_{00}}\, dt + \frac{g_{0\alpha} dx^{\alpha}}{c\sqrt{g_{00}}} = 0 \qquad (7)$$

as correct generalization for the Eqs.(5). From Eqs.(7) we obtain

$$\sqrt{g_{00}} + \frac{g_{0\alpha}}{c\sqrt{g_{00}}} \frac{dx^{\alpha}}{dt} = 0. \qquad (8)$$

## 2.Proper length revisited.

Note that under condition (8) interval $ds^2 = c^2 d\tau_{\text{ph}}^2 - dl_{\text{ph}}^2$, changes from the interval with spatial part only, i.e. $ds^2 = -dl_{\text{ph}}^2$. Suppose that on the time interval $t \in [t_1, t_2]$ global solution $\tilde{x}^{\alpha}(t)$ of the Eq.(8) exist and the corresponding boundary conditions: $\tilde{x}^{\alpha}(t_1) = \tilde{x}_1^{\alpha}(t_1), \tilde{x}^{\alpha}(t_2) = \tilde{x}_2^{\alpha}(t_2)$ is satisfied. Finally we obtain

$$l_{\text{ph}} = sgn(t_2 - t_1) \int_{t_1}^{t_2} \sqrt{\left(-g_{\alpha\beta}(\tilde{x}^{\alpha}(t), t) + \frac{g_{0\alpha}(\tilde{x}^{\alpha}(t), t) g_{0\beta}(\tilde{x}^{\alpha}(t), t)}{g_{00}(\tilde{x}^{\alpha}(t), t)}\right) \frac{d\tilde{x}^{\alpha}}{dt} \frac{d\tilde{x}^{\beta}}{dt}}\, dt.$$

$$\sqrt{g_{00}} + \frac{g_{0\alpha}}{c\sqrt{g_{00}}} \frac{d\tilde{x}^{\alpha}}{dt} = 0, \qquad (9)$$

$$\tilde{x}^{\alpha}(t_1) = \tilde{x}_1^{\alpha}(t_1), \tilde{x}^{\alpha}(t_2) = \tilde{x}_2^{\alpha}(t_2).$$

## 3.Example.J.S. Bell's problem.

Its gist consists in the following [3],[4]).Two rockets **B** and **A** are set in motion with constant proper acceleration so that the distance between them may remains constant and equal to the starting one $L_0$ from the viewpoint of an external observer **C**. One can simpler imaging here that the observer A operates the flight of the moving-away rockets and has a radar, by means of which he controls the constancy of the distance between them. From the viewpoint of an external observer **C** rockets **A** and **B** moving by laws $x_{\text{A}}(t)$ and $x_{\text{B}}(t)$ :

$$x_A(t) = \frac{c^2}{a}\left(\sqrt{1 + \frac{a^2 t^2}{c^2}} - 1\right),$$

$$x_B(t) = L_0 + \frac{c^2}{a}\left(\sqrt{1 + \frac{a^2 t^2}{c^2}} - 1\right).$$

(10)

In the case of motion with constant proper acceleration $a$ the interval $ds^2$ of the FR comuving both rockets takes the form (see [5] Eq.12.12):

$$ds^2 = \frac{c^2 dt^2}{1 + \frac{a^2 t^2}{c^2}} - \frac{2at\,dt\,dx}{\sqrt{1 + \frac{a^2 t^2}{c^2}}} - dx^2 - dy^2 - dz^2.$$

(11)

From Eqs.(8) one obtain

$$\frac{dx^1}{dt} = \frac{c}{\beta(t)\sqrt{1 + \beta^2(t)}}, \beta(t) = \frac{at}{c}.$$

(12)

Hence

$$x^1(t) = -\frac{c^2}{a}\ln\left|\frac{1 + \sqrt{1 + \beta^2(t)}}{\beta(t)}\right| + A.$$

(13)

Setting $t = t_1$ and take into account corresponding boundary condition $x^1(t)|_{t=t_1} = x^1(t_1) = 0$ one obtain:

$$\frac{c^2}{a}\ln\left|\frac{1 + \sqrt{1 + \beta^2(t_1)}}{\beta(t_1)}\right| = A(t_1).$$

(14)

Setting $t = t_2$ and take into account corresponding boundary condition $x^1(t_2) = L_0$ one obtain:

$$-\frac{c^2}{a} \ln \left| \frac{1 + \sqrt{1 + \beta^2(t_2)}}{\beta(t_2)} \right| + A(t_1) = L_0. \tag{15}$$

Substitution Eq.(14) into Eq.(15) gives

$$\frac{c^2}{a} \ln \left| \frac{\left(1 + \sqrt{1 + \beta^2(t_1)}\right) \beta(t_2)}{\left(1 + \sqrt{1 + \beta^2(t_2)}\right) \beta(t_1)} \right| = L_0. \tag{16}$$

From Eq. (16) by simple calculation (see [6] appendix Eqs.(C.3)-(C.6)) one obtain

$$\frac{\beta(t_2)}{\beta(t_1)} = \cosh\left(\frac{aL_0}{c^2}\right) + \sinh\left(\frac{aL_0}{c^2}\right) \times \sqrt{1 + \beta^2(t_1)}. \tag{17}$$

Substitution Eqs.(13),(17) into Eq.(9) gives

$$l_{\mathbf{ph}}(t) = sgn(t_2 - t_1) \frac{c^2}{a} \int\limits_{t_1=t}^{t_2(t)} \frac{dt}{t} =$$

$$= \frac{c^2}{a} \ln\left( \cosh\left(\frac{aL_0}{c^2}\right) + \sinh\left(\frac{aL_0}{c^2}\right) \times \sqrt{1 + \beta^2(t)} \right), \beta(t) = \frac{at}{c}. \tag{18}$$

Eq.(18) originally was obtained by S.A.Podosenov by the completely geometrical consideration [6].

## 4. Instant proper length measurement between J.S.Bell's rockets.

Instant proper length measurement between J.S.Bell's rockets in this subsection is considered successfully.

Let as consider J.S.Bell's problem in canonical parametrization such that $(c = 1)$ [3]:

$$x_B(\tau) = \frac{1}{a}[\cosh(a \cdot \tau) - 1] + x_B^0,$$

$$x_A(\tau) = \frac{1}{a}[\cosh(a \cdot \tau) - 1] + x_A^0,$$

$$x_B^0 - x_A^0 = L_0, \qquad (19)$$

$$\tau(t) = \frac{1}{a} \ln\left(a \cdot t + \sqrt{1 + a^2 \cdot t^2}\right),$$

$$t(\tau) = \frac{1}{a} \sinh(a \cdot \tau).$$

We set below $x_A^0 = 0, x_B^0 = L_0$. By using transformations (corresponding with the canonical parametrization (Eq.(19))

$$\rho = x - \frac{1}{a}[\cosh(a \cdot \tau(t)) - 1],$$

$$t(\tau) = \frac{1}{a} \sinh(a \cdot \tau). \qquad (20)$$

by the canonical way the interval $ds^2$ of the FR comuving both rockets takes the form [3]:

$$ds^2 = d\tau^2 - d\rho^2 - 2d\tau d\rho \sinh(a \cdot \tau) - dy^2 - dz^2.$$

$$g_{\tau\tau} = 1, g_{\rho\rho} = -1, g_{\tau\rho} = -\sinh(a \cdot \tau), g_{yy} = -1, g_{zz} = -1 \qquad (21)$$

By using Eqs.(1)-(3) we obtain

$$ds^2 = [d\tau - d\rho \sinh(a \cdot \tau)]^2 - d\rho^2 \cosh(a \cdot \tau) - dy^2 - dz^2. \tag{22}$$

Let us place between rockets **A** and **B** a continuum of dot rockets (Pic.1) and we shall consider two dot rockets $\mathbf{A}_i$ and $\mathbf{A}_{i+1}$ the initial distance between which is infinitesimal size $dL_0$.

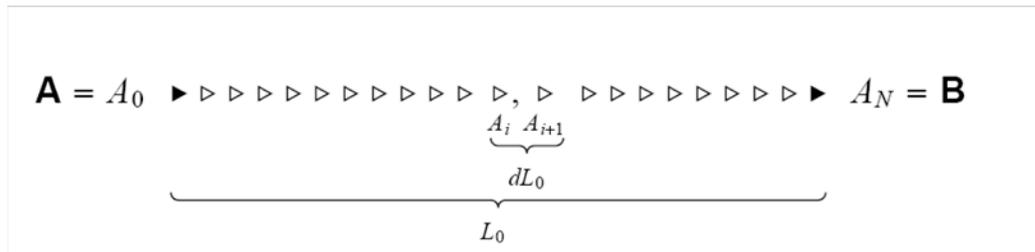

**Pic. 1.** Continuum of dot rockets between rockets **A** and **B** and two dot rockets $\mathbf{A}_i$ and $\mathbf{A}_{i+1}$ the initial distance between which is infinite smal size $dL_0$.

Let's assume, that clocks in rockets are synchronized, according to canonical procedure described in [1]. That in a considered case obviously physically realizable procedure.

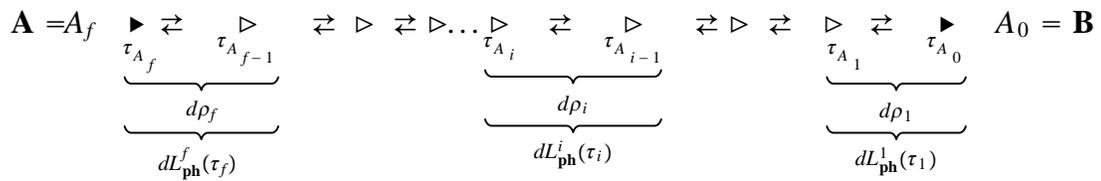

**Pic. 2.** Mental experiment explaining procedure of measurement **instant physical distance** between Bell's rockets.

Let us consider now infinite set of infinitely close physical events $\tau_{A_i}$, consisting that the observer sitting in rocket $A_i$ during the moment of proper time $\tau_i$ sends a light signal aside rockets $A_{i+1}$ (Pic.2.) which having reflected, comes back back in rocket $A_i$. Thus any two events $\tau_{A_i}$ and $\tau_{A_{i+1}}$ are assumed simultaneous in sense of physical time $\tau_{ph}$, i.e. $d\tau_{ph} = 0$, see [1]. Suppose also that $\tau_{i+1} - \tau_i = d\tau_i$. Hence by virtue of a condition of a simultaneity of any pair events $\{\tau_{A_i}, \tau_{A_{i+1}}\}$ we have the following equality:

$$d\tau_i = \sinh(a \cdot \tau_i) d\rho_i. \tag{23}$$

By using Eq.(2) and Eq.(23) one obtain

$$dL_{ph}^i(\tau_i) = \sqrt{\cosh^2(a \cdot \tau_i) d\rho_i^2(\tau_i)} =$$

$$= \left(\cosh(a \cdot \tau_i) \frac{d\rho_i(\tau_i)}{d\tau_i}\right) d\tau_i \tag{24}$$

By integration Eq.(24) we obtain

$$L_{ph}(\tau) = \sum_i dL_{ph}^i(\tau_i) =$$

$$= sgn(\tau_f - \tau_0) \int_{\tau_0}^{\tau_f(\tau_0)} \left(\cosh(a \cdot \tau) \frac{d\rho(\tau)}{d\tau}\right) d\tau =$$

$$= sgn(\tau_f - \tau_0) \int_{\tau_0}^{\tau_f(\tau_0)} \coth(a \cdot \tau) d\tau = \tag{25}$$

$$= sgn(\tau_f - \tau_0) \frac{1}{a} \ln\left|\frac{\sinh(a \cdot \tau_f(\tau_0))}{\sinh(a \cdot \tau_0)}\right|.$$

By integration Eq.(23) one obtain

$$\rho(\tau) = \frac{1}{a} \ln\left|\tanh\left(\frac{a \cdot \tau}{2}\right)\right| + A \tag{26}$$

Setting $\tau = \tau_f$ and take into account corresponding boundary condition $\rho(\tau)|_{\tau=\tau_f} = x_A^0 = 0$ one obtain:

$$\frac{1}{a} \ln\left|\tanh\left(\frac{a \cdot \tau_f(\tau_0)}{2}\right)\right| + A(\tau_f) = 0. \tag{27}$$

Setting $\tau = \tau_0$ and take into account corresponding boundary condition $\rho(\tau)|_{\tau=\tau_0} = x_B^0 = L_0$ one obtain:

$$\frac{1}{a} \ln\left|\tanh\left(\frac{a \cdot \tau_0}{2}\right)\right| + A(\tau_f) = L_0. \tag{28}$$

By subtracting Eq.(28) from Eq.(27) one obtain:

$$\frac{1}{a} \ln\left|\frac{\tanh\left(\frac{a \cdot \tau_f(\tau_0)}{2}\right)}{\tanh\left(\frac{a \cdot \tau_0}{2}\right)}\right| = -L_0. \tag{29}$$

From Eq.(29) we obtain

$$\frac{\tanh\left(\frac{a \cdot \tau_f(\tau_0)}{2}\right)}{\tanh\left(\frac{a \cdot \tau_0}{2}\right)} = \exp(-a \cdot L_0),$$

$$\tanh\left(\frac{a \cdot \tau_f(\tau_0)}{2}\right) = \left(\tanh\left(\frac{a \cdot \tau_0}{2}\right)\right) \cdot \exp(-a \cdot L_0). \tag{30}$$

Hence

$$\tau_f(\tau_0) = \frac{2}{a} \mathbf{Arth}\left[\left(\tanh\left(\frac{a \cdot \tau_0}{2}\right)\right) \cdot \exp(-a \cdot L_0)\right] \qquad (31)$$

By substitution Eq.(31) into Eq.(25) we obtain

$$L_{\mathbf{ph}}(\tau_0) = sgn(\tau_f - \tau_0)\frac{1}{a} \ln\left|\frac{\sinh(a \cdot \tau_f(\tau_0))}{\sinh(a \cdot \tau_0)}\right| =$$

$$= sgn(\tau_f - \tau_0)\frac{1}{a} \ln\left|\frac{\sinh\left[2 \cdot \mathbf{Arth}\left[\left(\tanh\left(\frac{a \cdot \tau_0}{2}\right)\right) \cdot \exp(-a \cdot L_0)\right]\right]}{\sinh(a \cdot \tau_0)}\right|. \qquad (32)$$

By using equality: $\sinh[2 \cdot \mathbf{Arth}(z)] = \frac{2z}{1-z^2}, (z^2 < 1)$ one obtain

$$\sinh\left[2 \cdot \mathbf{Arth}\left[\left(\tanh\left(\frac{a \cdot \tau_0}{2}\right)\right) \cdot \exp(-a \cdot L_0)\right]\right] =$$

$$= \frac{2\left(\tanh\left(\frac{a \cdot \tau_0}{2}\right)\right) \cdot \exp(-a \cdot L_0)}{1 - \left(\tanh^2\left(\frac{a \cdot \tau_0}{2}\right)\right) \cdot \exp(-2a \cdot L_0)}. \qquad (33)$$

By substitution Eq.(33) into Eq.(32) finally we obtain

$$L_{\mathbf{ph}}(\tau_0) = -\frac{1}{a} \ln\left[\frac{2\left(\tanh\left(\frac{a \cdot \tau_0}{2}\right)\right) \cdot \exp(-a \cdot L_0)}{\left(1 - \left(\tanh^2\left(\frac{a \cdot \tau_0}{2}\right)\right) \cdot \exp(-2a \cdot L_0)\right) \cdot \sinh(a \cdot \tau_0)}\right] \qquad (34)$$

Suppose that: $a \gg 1$ such that $a \cdot L_0 \gg 1, a \cdot \tau_0 \gg 1$. Thus from Eq.(34) one obtain

$$L_{\text{ph}}(\tau_0) \propto -\frac{1}{|a|} \ln\left[\frac{2 \cdot \exp(-a \cdot L_0)}{\sinh(a \cdot \tau_0)}\right] \propto$$

$$L_0 + \frac{1}{|a|} \ln[\sinh(|a| \cdot \tau_0)], \tau_0 \to \infty.$$

(35)

## 5. Rotating Frame with $\omega = $ const.

The line element for the cylindrical coordinate system of the Minkowski space are

$$ds^2 = dT^2 - dR^2 - R^2 d\Phi^2 - dZ^2. \quad (5.1)$$

Coordinate transformation, where upper case coordinates represent the inertial frame **K**, lower case denote the rotating frame **k**, and the axis of rotation is coincident with both the $Z$ and $z$ axes, are

$$T = t, R = r, \Phi = \phi + \omega t, Z = z. \quad (5.2)$$

Here $\omega$ is the angular velocity of the disk, and $t$ the coordinate time for the rotating system, is the proper time of a standard clock located at the origin of the rotating coordinate frame, i.e., it is equivalent to any standard clock at rest in $K$. Note that $t$ is only a coordinate. From (5.1)-(5.2) one obtains the metric of the coordinate grid in $k$:

$$ds^2 = (1 - r^2\omega^2)dt^2 - dr^2 - r^2 d\phi^2 - 2r^2\omega d\phi dt - dz^2 = g_{ij}dx^i dx^j. \quad (5.3)$$

where the covariant form of the metric $g_{ij}, i,j = 0,1,2,3$ are

$$g_{ij} = \begin{pmatrix} (1 - r^2\omega^2) & 0 & -r^2\omega & 0 \\ 0 & -1 & 0 & 0 \\ -r^2\omega & 0 & -r^2 & 0 \\ 0 & 0 & 0 & -1 \end{pmatrix}. \tag{5.4}$$

From Eq.(8) one obtain

$$\sqrt{1 - r^2\omega^2} - \frac{r^2\omega}{\sqrt{1 - r^2\omega^2}} \frac{d\varphi}{dt} = 0. \tag{5.5}$$

From Eq.(5.5) one obtain

$$\frac{d\varphi}{dt} = \frac{1 - r^2\omega^2}{r^2\omega}. \tag{5.6}$$

Thus

$$\varphi(t) = \frac{1 - r^2\omega^2}{r^2\omega} t + A. \tag{5.7}$$

Inserting into Eq.(5.8) $t = t_1$ and take into account corresponding boundary condition $\varphi(t)|_{t=t_1} = \varphi(t_1) = 0$ one obtain:

$$\frac{1 - r^2\omega^2}{r^2\omega} t_1 + A(t_1) = 0. \tag{5.8}$$

$$A(t_1) = -\frac{1 - r^2\omega^2}{r^2\omega} t_1.$$

Inserting into Eq.(5.8) $t = t_2$ and take into account corresponding boundary condition $\varphi(t)|_{t=t_2} = \varphi(t_2) = -2\pi$ one obtain:

$$\frac{1-r^2\omega^2}{r^2\omega}t_2 + A(t_1) = 2\pi. \tag{5.9}$$

Substitution Eq.(5.8) into Eq.(5.9) gives

$$\frac{1-r^2\omega^2}{r^2\omega}t_2 - \frac{1-r^2\omega^2}{r^2\omega}t_1 = 2\pi,$$

$$t_2 - t_1 = \frac{2\pi}{\varpi}, \varpi = \frac{1-r^2\omega^2}{r^2\omega}. \tag{5.10}$$

From Eq.(9) and Eq.(5.7)-Eq.(5.10) one obtain:

$$l_{\mathbf{ph}} = sgn(t_2 - t_1) \int_{t_1}^{t_2} \sqrt{\left(-g_{22} + \frac{g_{02}^2}{g_{00}}\right) \frac{d\varphi}{dt}\frac{d\varphi}{dt}} \, dt =$$

$$\int_{t_1}^{t_2} \sqrt{\left[r^2 + \frac{r^4\omega^2}{(1-r^2\omega^2)}\right]\left(\frac{1-r^2\omega^2}{r^2\omega}\right)^2} \, dt =$$

$$\int_{t_1}^{t_2} dt \sqrt{\frac{r^2 - r^4\omega^2 + r^4\omega^2}{1-r^2\omega^2}} \left(\frac{1-r^2\omega^2}{r^2\omega}\right) = \int_{t_1}^{t_2} dt \frac{\sqrt{1-r^2\omega^2}}{r^2\omega} = \tag{5.11}$$

$$\frac{2\pi r^2\omega}{1-r^2\omega^2} \frac{\sqrt{1-r^2\omega^2}}{r^2\omega} = \frac{2\pi}{\sqrt{1-r^2\omega^2}}.$$

# References

[1] Landay L.D. and Lifshitz E.M. The Classical Theory of Fields.

 Moscow, 1962.
[2] Zelmanov A. On the Relativistic Theory of an Anisotropic Inhomogeneous Universe. The journal for General Relativity, gravitation and cosmology. Vol.1, 2008.
[3] Gershtein S.S., Logunov A. A. J.S. Bell's problem. Phys. Part. Nucl. Volume 29, Issue 5, pp. 463-468
[4] Bell J.S. Speakable and Unspeakable in Quantum Mechanics (Cambridge University Press, Cambridge, 1993), p.67.
[5] Logunov A. A., Lectures in Relativity and Gravitation: A Modern Outlook Pergamon Press, Oxford, 1990.
[6] Foukzon J., Podosenov S.A. Generalized Principle of limiting 4-dimensional symmetry. http://arxiv.org/abs/0805.1644